\documentclass[a4paper,12pt]{article}      

\usepackage{fullpage}
\usepackage{titlesec}

\usepackage{amsmath}
\usepackage{amsfonts}               
\usepackage{hyperref}		

\usepackage{color}

\usepackage{graphicx}		
\usepackage{epstopdf}		

\usepackage{empheq}
\newcommand*\widefbox[1]{\fbox{\hspace{2em}#1\hspace{2em}}}

\numberwithin{equation}{section}

\newcommand{\e}{\eta} 		
\newcommand{\Omg}{\Omega} 	

\newcommand{\di}{\mathrm{d}} 

\newcommand{\wtd}[1]{\widetilde{#1}} 	
\newcommand{\ba}[1]{\overline{#1}}		

\newcommand{\sk}[1]{ |#1] } 			
\newcommand{\ak}[1]{ |#1\rangle }		
\renewcommand{\sb}[1]{ [#1| }			
\newcommand{\ab}[1]{ \langle#1| }		

\newcommand{\SB}[1]{ [#1] }					
\newcommand{\AB}[1]{ \langle #1 \rangle }	
\newcommand{\ASB}[1]{ \langle #1 ] }		
\newcommand{\SAB}[1]{ [ #1 \rangle }		

\newcommand{\h}[1]{\hat{#1}} 		
\newcommand{\ih}{\hat{i}} 			
\newcommand{\jh}{\hat{j}} 			
\newcommand{\Ph}{\hat{P}} 			

\newcommand{\MHV}{\mathrm{MHV}} 		
\newcommand{\NkMHV}{\mathrm{N^{k}MHV}} 	

\newcommand{\M}{\mathcal{M}} 						
\newcommand{\N}{\mathcal{N}} 						
\newcommand{\Order}[1]{\mathcal{O}\left(#1\right)} 	
\newcommand{\goesto}{\longrightarrow}			
\newcommand{\ifac}[1]{ \frac{1}{#1!} }			
\newcommand{\nonum}{\nonumber} 					

\titleformat*{\section}{\large\bf}
\titleformat*{\subsection}{\normalsize\bf}

\begin{document}

\title{\large\bf Bonus scaling and BCFW in $\N = 7$ supergravity}
\author
{
	\normalsize Jin-Yu Liu and En Shih \\
	\small\it Department of Physics, National Taiwan University, Taipei 10617, Taiwan \\
	\small \emph{Email:} \texttt
		{
		\href{mailto:k5438777@gmail.com}{k5438777@gmail.com},
		\href{mailto:seanstone5923@gmail.com}{seanstone5923@gmail.com}
		}
}
\date{\small \it November 20, 2014}
\maketitle


\begin{abstract} 
In search of natural building blocks for supergravity amplitudes, a tentative criteria is term-by-term bonus $z^{-2}$ large momentum scaling. For a given choice of deformation legs, we present such an expansion in the form of a BCFW representation in $\N=7$ supergravity based on a special shift. We will show that this improved scaling behavior, with respect to the fully $\N=8$ representation, is due to its automatic incorporation of the so called bonus relations.
\end{abstract}


\section{Introduction}
One of the fascinating themes in the study of planar $\N=4$ SYM, is that the amplitude is often a solution to a geometric question. The now famous example is  the realization that the building blocks for the $n$-point $\N=4$ SYM amplitude with $k$- negative helicity gluons, constructed via the Britto, Cachazo, Feng and Witten (BCFW) recursion relation~\cite{BCFW}, are associated with positive cells of a Grassmannian G(k,n)~\cite{NimaGrass, NimaBigBook}, the moduli space of $k$-planes in $n$-dimensional space. 
 
A natural question is whether such structure exists outside of $\N=4$ SYM. Certain progress has been made for $\N=6$ super Chern-Simons matter theory (CSM)~\cite{ABJM1, ABJM2}, in the context of an orthogonal Grassmannian~\cite{LeeOG, HW, HWX}. The common property between $\N=4$ SYM and $\N=6$ CSM theory is that both allow for color decomposition such that color ordered amplitudes can be defined, and the theories enjoy an infinite dimensional Yangian symmetry~\cite{Henn}. In fact the building blocks that arise from the recursion are individually Yangian invariant.    

Both of the above properties are absent in gravity, and thus it may be unclear how to proceed. Instead we can ask, if there are natural building blocks for gravity amplitudes, what would be the nice property one can ask from it, similar to Yangian invariance for the gauge theories. One special property of gravity amplitudes is the asymptotic behavior in the large momentum limit. Indeed it was known that in the BCFW recursion, if one shifts $\ak{i}$ and $\sk{j}$, where $i$ and $j$ are a positive and negative helicity graviton respectively, as the deformation parameter $z$ is taken to infinity, the amplitude behaves as $1/z^2$~\cite{NimaKaplan}.\footnote{ Recently, it has been shown that this asymptotic behavior can be attributed to the permutation invariance of gravity amplitudes~\cite{David}.} This is to be compared with $1/z$ of Yang-Mills.      

Thus we propose that a criteria for a ``good" building block is good large-$z$ scaling under any pair of shifted momenta. Note that in a generic BCFW representation, individual terms can behave as $1/z$ and only cancel in the sum. To begin, we will relax our criteria and ask: if one chooses two particular legs to deform, is there a representation such that individual terms scale as $1/z^2$ under large deformation? We will show that indeed such a representation exists, in the form of a BCFW recursion in $\N=7$ supergravity, constructed out of a ``bad-shift". $\N=7$ supergravity has the same on-shell degrees of freedom as with $\N=8$ supergravity, only with a reduced set of supersymmetry being manifest. However the reduced symmetry allows us to exploit the $1/z^2$ fall off of the full $\N=8$ amplitude. More precisely we claim that if one constructs the $\N=7$ amplitude under the following $\SAB{j^+,i^-}$ bad shift:
\begin{equation}
	\label{key}
\sk{j^+} \rightarrow \sk{j^+}+w\,\sk{i^-} ,\quad
\ak{i^-} \rightarrow \ak{i^-}-w\,\ak{j^+} ,\quad
\e_{j^+} \rightarrow \e_{j^+}+w\,\e_{i^-} .
\end{equation}
Then the individual terms in the BCFW expansion scale at large $z$ as $1/z^2$ under the following $\SAB{i^-,j^+}$ shift of the same primary shifted legs:
\begin{equation}
\sk{i^-} \rightarrow \sk{i^-}+z\,\sk{j^+} ,\quad
\ak{j^+} \rightarrow \ak{j^+}-z\,\ak{i^-} ,\quad
\e_{i^-} \rightarrow \e_{i^-}+z\,\e_{j^+} .
\end{equation}
Note that the bad-shift in $\N$-supergravity behaves as $z^{8-\N}/z^2=z^{6-\N}$, and thus it has sufficient fall off for a valid recursion relation for $\N=7,8$. As we will argue, the reason why $\N=7$ bad-shift recursion allows for term by term $1/z^2$ fall off is because it secretly uses the $1/z^2$ fall off of the full amplitude. For a valid BCFW representation, all one needs is that the amplitude vanish as $z\rightarrow\infty$, thus even though gravity amplitudes behave as $1/z^2$, the usual BCFW recursion is blind to such improved fall off. On the other hand, for the $\N=7$ bad-shift, the large-$z$ fall off behaves as $1/z$ precisely because of the $1/z^2$ of the full $\N=8$ amplitude. Thus, the $1/z$ fall off is crucial for the validity of the $\N=7$ bad shift. The presence of $1/z^2$ fall off implies extra ``bonus relations" for individual BCFW terms~\cite{NimaSimplest}. As we will show, for MHV amplitudes, it is precisely due to these bonus relations that the $\N=7$ bad shift exhibit improved fall off relative to $\N=8$.

Note that representations with term by term $1/z^2$ fall off are already known for MHV amplitudes~\cite{Nguyen:2009jk}. However, no known expression with such properties exist beyond the MHV sector. The $\N=7$ bad shift allows for such a representation beyond MHV level. This special property of the $\N=7$ bad-shift has already been noted at the six-point level in Hodges work~\cite{Hodges:2011wm}. In this paper we present a proof extending to general tree-level amplitudes.

This paper is organized as follows: first we introduce BCFW recursion in the formalism of $\N=7$ supergravity, and examine its validity under different scenarios, leading us to investigate the large $z$ behavior of the $\SAB{+,-}$ ``bad shift" representation. We then present a proof for term-by-term $\Order{z^{-2}}$ scaling of the ``bad shift" representation under a correspondingly chosen test shift. Furthermore, we discover the improved scaling in $\N=7$ is related to bonus relations in $\N=8$.


\section{$\N=7$ superamplitudes}

Here we review the derivation of $\N=7$ supergravity amplitudes from its $\N=8$ counterpart, as well as its large $z$ behavior. This discussion follows~\cite{Elvang:2011fx}.

\subsection{From $\N=8$ to $\N=7$}

We formulate $\N=8$ supergravity using a on-shell superspace by introducing eight Grassmann variables $\eta^A$, labeled by the SU(8) index $A=1...8$. This allows us to associate the states of various helicites in the $\N=8$ theory with components of different orders of $\eta$ in an on-shell chiral superfield, which we write as
\begin{align}
	\nonum \Omg &=
	h^+ 
	+\psi_A\e^A 
	+\ifac{2}v_{AB}\e^A\e^B 
	+\ifac{3}\chi_{ABC}\e^A\e^B\e^C 
	+\ifac{4}S_{ABCD}\e^A\e^B\e^C\e^D
	\\&\quad
	+\ifac{3}\chi^{ABC}\e^5_{ABC} 
	+\ifac{2}v^{AB}\e^6_{AB}
	+\psi^A\e^7_A
	+h_-\e^8 .
\end{align}
where $\e^5_{ABC} \equiv \ifac{5}\epsilon_{ABCDEFGH}\e^D\e^E\e^F\e^G\e^H$, and other $\e$ polynomials are similarly defined.

When we reduce the manifest supersymmetry from $\N=8$ to $\N=7$, the on-shell states separate into two superfields, which are obtained respectively from two different ways of reducing supersymmetry: setting $\e^8$ to zero or integrating away $\e^8$.
\begin{subequations}
\begin{align}
	\Phi^+ &\equiv \Omg|_{\e^8\rightarrow 0}
	= \int \di\e^8 \ \e^8 \, \Omg \,, \\
	\Phi^- &\equiv \int\di\e^8 \ \Omg \,.
\end{align}
\end{subequations}
The explicit forms of the superfields are:
\begin{subequations}
\begin{align}
\nonum \Phi^+ &= 
h^+
+\psi_A\e^A
+\ifac{2}v_{AB}\e^A\e^B 
+\ifac{3}\chi_{ABC}\e^A\e^B\e^C 
+\ifac{3}S^{8ABC}\e^4_{ABC}
\\&\quad
+\ifac{2}\chi^{8AB}\e^5_{AB}
+v^{8A}\e^6_{A}
+\psi^8\e^7 ,
\\
\nonum \Phi^-
&= \psi_8
+v_{8A}\e^A
+\ifac{2}\chi_{8AB}\e^A\e^B
+\ifac{3}S_{8ABC}\e^A\e^B\e^C
+\ifac{3}\chi^{ABC}\e^4_{ABC}
\\ &\quad
+\ifac{2}v^{AB}\e^5_{AB}
+\psi^{A}\e^6_{A}
+ h^-\e^7 .
\end{align}
\label{N=7_multiplets}
\end{subequations}
The indices are now summed from 1 to 7, and $\e^4_{ABC}\equiv\ifac{4}\epsilon_{ABCDEFG}\e^D\e^E\e^F\e^G$. Note that setting $\e^8$ to zero can be represented by a integration over $\e^8$ after multiplying by $\e^8$. The $\Phi^+$ multiplet has helicty +2, and contains the positve helicity graviton $h^+$, while $\Phi^-$ has helicty +3/2, and contains the negative helicity graviton $h^-$. We will use a $+$ sign to mark quantities associated with the $\Phi^+$ multiplet, while quantities associated with the $\Phi^-$ multiplet will be marked with a $-$ sign.

Using the same operations, $\N=7$ amplitudes can be derived from the corresponding $\N=8$ amplitudes. As an example, the $\N=7$ MHV 3-point graviton scattering amplitude is obtained from the $\N=8$ MHV 3-point amplitude as follows:
\begin{equation}
	\M_3(1^-2^-3^+) = \int \di\e_1^8\di\e_2^8\di\e_3^8 \ \e_3^8 \, \M_3^{\MHV}(123) .
\end{equation}
Here the first subscript of $\eta$ refers to the associated particle number, while the superscript refers to the SU(8) index.

For a general $\NkMHV$ amplitude, there will be $k+2$ external legs in the $\Phi^-$ multiplet, which we denote by the set $\{x\}$, and $n-k-2$ external legs in the $\Phi^+$ multiplet, which we denote by the set $\{y\}$. Then we have the following map between $\N=7$ and $\N=8$ amplitudes:
\begin{equation}
\M^{\N=7}(\{x\},\{y\}) = \int \left[\prod_{a=1}^n \di\e_a^8\right] \left[\prod_{b \in \{y\}}\eta_b^8\right]\M^{\N=8} .
\label{map} 
\end{equation}
Or more explicitly,
\begin{equation}
\M^{\N=7}_n (1^-,\cdots,(k+2)^-,(k+3)^+,\cdots,n^+) = \int \di\e_1^8\cdots\di\e_n^8 \ \e_{k+3}^8\cdots\eta_n^8 \, \M_n^{\N=8}(1,\cdots, n).
\end{equation}


\subsection{BCFW in the $\N=7$ formalism}

Validity of a BCFW representation requires the amplitude vanish as the deformation parameter $z$ goes to infinity:

\begin{equation}
\nonum \sk{\ih} = \sk{i} + z\sk{j} , \quad
\ak{\jh} = \ak{j} + z\ak{i} , \quad
\h{\e_i} = \e_i + z \, \e_j ,
\end{equation} 
\begin{equation}
\M(z) \goesto 0 \ \ \ \text{as} \ \ \ z \goesto \infty \,.
\end{equation}

$\N=8$ amplitudes scale as $\Order{z^{-2}}$ for large $z$. In the case of $\N=7$, we can deduce the large $z$ behavior by relating the $\N=7$ amplitude to the parent $\N=8$ using (\ref{map}). Unlike in the case of $\N=8$, amplitudes in $\N=7$ specialize into different supermultiplet configurations for lines $i,j$ which may show different large $z$ behavior.

Note that in order to deduce the large $z$ behavior of $\N=7$ from $\N=8$ using (\ref{map}), we need to take into the subtlety that for $\N=8$, we shift $\h{\e}_i^A$ for $A=1...8$, while for $\N=7$, we only shift for $A=1...7$. Thus we need to somehow unshift $\h{\e}_i^8$. This can easily be done by a change of variables. We define 
\begin{equation}
\e_i^8 = \ba{\e}_i^8 - z\ba{\e}_j^8, \quad 
\e_a^8 = \ba{\e}_a^8 \quad \text{for}\ a \neq i.
\label{change}
\end{equation}
The Jacobian is simply 1. Now we can promote (\ref{map}) into a relation for the shifted variables:
\begin{equation}
\M^{\N=7}(z) = \int \left[\prod_{a=1}^n \di\ba{\e}_a^8\right] \left[\prod_{b \in \{y\}}\e_b^8(\ba{\e}_c^8)\right] \M^{\N=8}(z) \,,
\end{equation}
where $\e_b^8$ is a function of $\ba{\e}_c^8$, as defined by (\ref{change}).\\

We can now analyze different scenarios for which mutiplet the lines $i,j$ in our $\SAB{i,j}$ shift sits in: 
\begin{itemize}
	\item For $\SAB{i^-,j^+}$ and $\SAB{i^-,j^-}$: Since $i$ is not in the $\Phi^+$ multiplet, $\eta_b^8$ does not contain any $z$ dependence, and hence the $\N=7$ amplitude behaves as $\Order{z^{-2}}$ at large $z$ exactly like $\N=8$. 
 	
	\item For $\SAB{i^+,j^+}$: Now $i$ belongs to the $\Phi^+$ multiplet, so naively applying a change of variable, one would pick up a $z$ factor. However the $z$ will be proportional to $\ba{\e}_j$ which is already present in $\e_b^8$ and thus this term drops out, i.e. $(\ba{\e}_i-z\ba{\e}_j)\ba{\e}_j=\ba{\e}_i\ba{\e}_j$. Thus we see for this shift, the $\N=7$ amplitude again behaves as $\Order{z^{-2}}$ at large $z$ exactly like $\N=8$.
 	
	\item For $\SAB{i^+,j^-}$:  Now $i$ belongs to the $\Phi^+$ multiplet, while $j$ does not, so $\e_b^8$ obtains an overall factor of $z$. Thus the large $z$ behavior for $\N=7$ amplitude behaves as $\Order{z} \times \Order{z^{-2}} = \Order{z^{-1}}$.
\end{itemize}

From the above we conclude that for the ``good" shifts $\SAB{i^-,j^+}$, $\SAB{i^-,j^-}$, $\SAB{i^+,j^+}$, the $\N=7$ amplitude behaves as $1/z^2$ just as the $\N=8$ parent. The BCFW built for $\N=7$ from the good shifts will be using the same $1/z$ pole as the $\N=8$ parent. Thus the BCFW built from the $\SAB{+,-}$ "bad" shift in $\N=7$ is secretly using information of the $1/z^2$ behavior of the $\N=8$ amplitude. In the following section, we will demonstrate that the $\N=7$ BCFW expansion built from the $\SAB{j^+,i^-}$ ``bad shift" indeed has bonus behavior in the form of term-by-term $\Order{z^{-2}}$ large-$z$ scaling under the $\SAB{i^-,j^+}$ test shift.


\section{Bonus $z$ scaling of $\N=7$ ``bad shift" BCFW terms}

\subsection{A particular $\SAB{-,+}$ test shift: $\NkMHV$ amplitudes}

We would like to prove that the $\N=7$ $\SAB{j^+,i^-}$ ``bad shift" BCFW terms have $\Order{z^{-2}}$ large $z$ fall off under the secondary $\SAB{i^-,j^+}$ test shift. Note our analysis can be easily applied to other helicity configurations as well, where the $\Order{z^{-2}}$ fall off is no longer present. Therefore, we start without fixing which superfields particles $i$ and $j$ belong to and construct the $\SAB{j,i}$ BCFW representation of the amplitude:
\begin{equation} 
\M_n(1,\cdots, i,\cdots, j,\cdots, n)
= \sum \int\di^7\e_{\Ph}  \ \M_L(-\Ph,\jh,\cdots) \, \frac{1}{P^2} \, \M_R(\Ph,\ih,\cdots) \  |_{\Ph^2=0} \,,
\label{bcfw}
\end{equation} 
\begin{equation} 
\sk{\jh} = \sk{j} + w \sk{i} ,\quad
\ak{\ih} = \ak{i} - w \ak{j} ,\quad
\h{\e}_j = \e_j + w \, \e_i \,.
\label{primary_shift}
\end{equation}
\begin{figure}[h]
	\centering
	\includegraphics[width=0.5\textwidth]{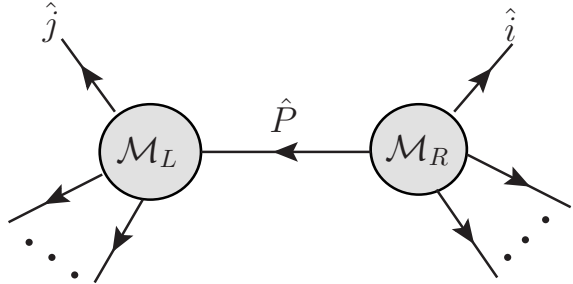}
	\caption{Diagram of a BCFW term}
	\label{fig:bcfw_term}
\end{figure}

For the on-shell condition $\Ph^2 = (P + w\,\sk{i}\ab{j})^2 = 0$, we can solve for $w$ and $\Ph$ in terms of $i,j$ and $P$. Leaving details of derivation to the appendix, the result is\footnote{ We adopt the ``mostly minus" metric convention, such that $p_k=\sk{k}\ab{k}$ and $s_{ij}=(p_i+p_j)^2=\SB{ij}\AB{ji}$ for massless particles.}.
\begin{equation} 
w = -\frac{P^2}{\ASB{j|P|i}} \,,
\end{equation}
\begin{equation} 
\Ph = \frac{P\ak{j}\sb{i}P}{\ASB{j|P|i}} \,.
\end{equation}

Let us now deform (\ref{bcfw}) by a $\SAB{i,j}$ test shift:
\begin{equation} 
\label{test_shift}
\sk{i}(z) = \sk{i} + z \sk{j} ,\quad
\ak{j}(z) = \ak{j} - z \ak{i} ,\quad
\e_i(z) = \e_i + z \, \e_j .
\end{equation}
Under the test shift, the amplitude is deformed into
\begin{align} 
\M_n(z) &= \sum \int d^7\e_{\Ph} \ \M_L(-\Ph(z),\jh(z),\cdots) \, \frac{1}{P^2(z)} \, \M_R(\Ph(z),\ih(z),\cdots)
\end{align}
Now $\sk{i},\ak{j},\e_i,P^2,\sk{\jh},\ak{\ih},\h{\e}_j,\sk{\Ph},\ak{\Ph}$ have become functions of $z$. Since the BCFW terms must have zero little group weight in $\Ph$, the $z$ dependence of the BCFW terms only comes from $\sk{i},\ak{j},\e_i,P^2,\sk{\jh},\ak{\ih},\h{\e}_j,\Ph$. By analyzing their large $z$ behavior individually, we can deduce the large $z$ behavior of the BCFW term as a whole. We thus proceed to do so.

From the $\SAB{i,j}$ test shift (\ref{test_shift}), deriving the large-$z$ behavior of $\sk{i},\ak{j},\eta_i,P^2$ is straightforward:
\begin{equation}
	\label{test_shifted_largez}
	\sk{i}(z) \goesto \Order{z} ,\quad
	\ak{j}(z) \goesto \Order{z} ,\quad
	\e_i(z) \goesto \Order{z} .
\end{equation}
\begin{equation}
	P^2(z) = P^2 - z\ASB{i|P|j} \goesto \Order{z} ,
\end{equation}
The primary deformed quantities $\sk{\jh},\ak{\ih},\h{\e}_j,\Ph$ transform under the test shift as
\begin{equation}
\label{primary_test_shifted}
\sk{\jh}(z) = \sk{j} + w(z) \sk{i}(z) ,\quad
\ak{\ih}(z) = \ak{i} - w(z) \ak{j}(z) ,\quad
\h{\e_j}(z) = \e_j + w(z) \e_i(z) ,
\end{equation}
\begin{align}
\Ph(z)
&= \frac{ -(P-p_j)\ak{i}\sb{j}(P+p_i) } { \ASB{i|P|j} } + \Order{z^{-1}} \goesto \Order{z^0} \,.
\end{align}
To determine the large-$z$ behavior of $\sk{\jh},\ak{\ih},\h{\e}_j$, we solve for the $z$-deformed primary shift parameter $w(z)$, and
expand it in powers of $z$:
\begin{align}
w(z) &= -\frac{1}{z} + \frac{ -P^2 - \ASB{j|P|j} + \ASB{i|(P-p_j)|i} } {\ASB{i|P|j}}\frac{1}{z^2}  + \Order{z^{-3}} .
\label{w_largez}
\end{align}
We expand to $\Order{z^{-2}}$ since the leading term gets canceled when we plug in expressions (\ref{test_shift}) and (\ref{w_largez}) into (\ref{primary_test_shifted}). We get:
\begin{align}
\label{primary_test_shifted_}
\sk{\jh}(z) &= \left(-\sk{i} + \frac{-P^2 - \ASB{j|P|j} + \ASB{i|(P-p_j)|i}}{\ASB{i|P|j}}\sk{j} \right) \frac{1}{z} + \Order{z^{-2}} , \nonum \\
\nonum \ak{\ih}(z) &= \left(\ak{j} + \frac{-P^2 - \ASB{j|P|j} + \ASB{i|(P-p_j)|i}}{\ASB{i|P|j}}\ak{i} \right) \frac{1}{z} + \Order{z^{-2}} ,\\
\h{\e_j}(z) &= \left(-\e_i + \frac{-P^2 - \ASB{j|P|j} + \ASB{i|(P-p_j)|i}}{\ASB{i|P|j}}\e_j \right) \frac{1}{z} + \Order{z^{-2}} \,.
\end{align}
Now we can read off their large-$z$ behavior. The results are organized below:
\begin{empheq}[box=\widefbox]{align}
 \label{largez}
    \nonum
    \\
    \nonum
 	\sk{i}(z) \goesto \Order{z} ,\quad
 	\ak{j}(z) &\goesto \Order{z} ,\quad
 	\e_i(z) \goesto \Order{z} \,,
 	\\
 	\nonum
    \sk{\jh}(z) \goesto \Order{z^{-1}} ,\quad
    \ak{\ih}(z) &\goesto \Order{z^{-1}} ,\quad
    \h{\e_j}(z) \goesto \Order{z^{-1}} \,,
    \\
    \nonum
    \Ph(z) &\goesto \Order{z^0} \,,
    \\
 	P^2(z) &\goesto \Order{z} .
\end{empheq}

With the large-$z$ scaling of $\sk{i},\ak{j},\e_i,P^2,\sk{\jh},\ak{\ih},\h{\e}_j,\Ph$ in hand, we can know how the BCFW term behaves at large $z$ by counting the orders of these contributing components. From (\ref{largez}) we see that $\sk{i},\e_i$, which have helicity $1/2$, behave as $\Order{z}$. On the other hand, $\ak{\ih}$, which has helicity $-1/2$, scales oppositely as $\Order{z^{-1}}$. We can write a general Ansatz that if particle $i$ contributes to the amplitude in the form of $\sk{i}^a\e_{i}^b\ak{\ih}^c$, then it scales as $\Order{z^{a+b-c}}$. 

In general, determining the orders of the spinors and the Grassmann variable can be nontrivial. However, in this case little group scaling of external leg $i$ trivializes the counting by fixing $a+b-c=2h_i$, where $h_i$ and $h_j$ are the helicities of the superfield corresponding to legs $i$ and $j$. Therefore, particle $i$ contributes $\Order{z^{2h_i}}$ at large $z$. A similar analysis shows that particle $j$ contributes $\Order{z^{-2h_j}}$ at large $z$. Since $\Ph$ approaches a constant at $z\rightarrow\infty$, the large $z$ scaling of each BCFW term is of:
\begin{empheq}[box=\widefbox]{equation}
 \Order{z^{2(h_i-h_j)-1}} .
 \label{scaling}
\end{empheq}

 Crucial to this result is the choice of the $\SAB{j,i}$ primary shift followed by $\SAB{i,j}$ test shift, which enjoys the cancellation of order $z^0$ terms while obtaining (\ref{primary_test_shifted_}) and thus ensures that the square spinors and the Grassmann variable scale oppositely to the angle spinors. Other choices would not have allowed us to determine the large $z$ scaling from the helicities alone. For example, if we chose a $\SAB{j,i}$ primary shift followed by a $\SAB{k,j}$ test shift, where $i \neq k$, then $\sk{k}$ and $\e_k$ would scale as $\Order{z}$ while $\ak{k}$ scale as $\Order{z^0}$. If particle $k$ contributes to the amplitude in the form of $\sk{k}^a\e_{k}^b\ak{k}^c$, then it would scale as $\Order{z^{a+b}}$, so $a+b-c=2h_k$ would not be sufficient to determine the large $z$ scaling contributed by particle $k$.

Note that up until this point the we have not designated the helicities of superfields $i$ and $j$. If we choose a $\SAB{j^+,i^-}$ ``bad" $\N$ supershift for supergravity, $h_j$ and $h_i$ would be separated by $\frac{8-\N}{2}$, such that the large $z$ scaling of each BCFW term be:
\begin{empheq}[box=\widefbox]{equation}
 \Order{z^{\N-9}} .
 \label{scaling2}
\end{empheq}
 We now specialize to the $\N=7$ $\SAB{j^+,i^-}$ ``bad shift" BCFW expansion under the secondary $\SAB{i^-,j^+}$ test shift. From the expressions for the $\N=7$ superfields (\ref{N=7_multiplets}), superfield $i$ has helicity $+3/2$ and therefore contributes $\Order{z^3}$ at large $z$, while superfield $j$ has helicity $+2$ and gives us $\Order{z^{-4}}$. $1/P^2$ gives $\Order{z^{-1}}$. Collectively, we find that the large $z$ scaling for the BCFW term is of:
\begin{empheq}[box=\widefbox]{equation}
\Order{z^{-2}} .
\end{empheq} 

We are lead to this result only if we specialize to the case where the $\SAB{j^+,i^-}$ bad shift is the primary shift. Other choices can result in $\Order{z^{-1}}$ or worse fall off. However, note that our counting is only indicative of the worst behavior, so the terms can actually have better fall off than shown by the counting. For example, both $\N=7$ $\SAB{j^+,i^+}$ and $\SAB{j^-,i^-}$ count to $\Order{z^{-1}}$, but explicit calculations have shown that some but not all of their BCFW terms behave as $\Order{z^{-2}}$.

Finally, note that the place where $\N=7$ plays a crucial role is the fact that the bad shift BCFW recursion is not valid for $\N<7$, while $\N=8$ does not distinguish between different shifts.


\subsection{General $\SAB{-,+}$ test shifts: the MHV case}

The above result fails for general BCFW test shifts other than the $\SAB{i^-,j^+}$ shift, and an alternative analysis is required. In general, there are many combinations of test shifts that we can choose from, however we are mainly concerned with the $\SAB{-,+}$ test shift, since it is the most relevant in the high energy limit. In the following, we analyze the large $z$ scaling under general $\SAB{-,+}$ test shifts in the MHV case.

\begin{figure}[h]
	\centering
	\includegraphics[width=0.5\textwidth]{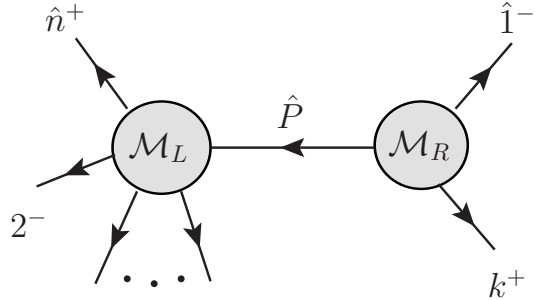}
	\caption{Diagram of a MHV ``bad shift" BCFW term}
\end{figure}

Choosing the $\SAB{n^+,1^-}$ primary shift, the amplitude factorizes into a $n-1$ point MHV subamplitude and a 3-point $\ba{\MHV}$ subamplitude. Similar to our previous analysis, first we solve for $w$ and $\Ph$:
\begin{equation}
w = \frac{\AB{1 k}}{\AB{n k}} \,,
\end{equation}
\begin{equation}
\label{MHV_P_hat}
\h{P} = -\left( \sk{k} + \frac{\AB{n 1}}{\AB{n k}}\sk{1} \right) \ab{k} .
\end{equation}

We now analyze the large $z$ scaling under different $\SAB{-,+}$ test shifts:

\begin{itemize}
\item For the $\SAB{1^-,n^+}$ shift: The proof in the previous section applies, and there is $\Order{z^{-2}}$ term by term behavior.

\item For the $\SAB{2^-,n^+}$ shift: There is $\Order{z^{-2}}$ term by term behavior. The large $z$ behavior of the deformed quantities are:
\begin{align}
\nonumber \Ph &\goesto \Order{z^0} \\
\nonumber \sk{2}(z) &= \sk{2} + z \sk{n} \\
\nonumber \ak{n}(z) &= \ak{n} - z \ak{2} \\
\nonumber \sk{\hat{n}}(z) &= \sk{n} + w \sk{1} \goesto \sk{n} \\
\ak{\hat{1}}(z) &= \ak{1} - w (\ak{n} - z\ak{2}) \goesto \Order{z^0} .\label{sh}
\end{align}
 In the large $z$ limit, dependence on $z$ only comes from the $n-1$ point subamplitude $\M_L$, also we see that $\sk{\hat{n}} \rightarrow \sk{n}$. Therefore, the chosen test shift is precisely a BCFW shift on the subamplitude $\M_L$ at large $z$, so the BCFW term must scale as $\Order{z^{-2}}$. 

\item For a $\SAB{2^-,m^+}$ shift (where $m \neq n$): Individual terms scale as $\Order{z^{-2}}$. The same argument as above applies if $m$ is not on the 3 point amplitude, so terms scale as $\Order{1/z^2}$. Moreover, the BCFW expansion is summed over all possible permutations, but there is only one diagram where $m$ is on the 3 point amplitude, therefore this term must also scale as $\Order{z^{-2}}$, since the existence of an $\Order{z^{-1}}$ part cannot be canceled by other terms.

\item For a $\SAB{1^-,m^+}$ test shift: The above argument fails and there are terms which do not behave as $\Order{z^{-2}}$.
\end{itemize}

 Summarizing the results above, we have demonstrated that for the MHV case, the $\N=7$ $\SAB{n^+,1^-}$ bad shift BCFW representation has $\Order{z^{-2}}$ term by term large $z$ scaling under $\SAB{1^-,n^+}$, $\SAB{2^-,n^+}$ and $\SAB{2^-,m^+}$ test shifts.


\subsection{Comparison to other formulas for supergravity amplitudes}

The large $z$ scaling of the ``bad shift" BCFW representation can be compared with the tree formula for MHV amplitudes by Nguyen, Spradlin, Volovich, and Wen~\cite{Nguyen:2009jk},  which also manifest $\Order{z^{-2}}$ large $z$ fall off term-by-term under certain test shifts. The formula chooses two legs as special, and involves a sum of terms each represented by a tree diagram. By directly counting the orders of $z$ in the $z$ deformed formula, we see that if at least one of test shift legs are special, then the term will scale as $\Order{z^{-2}}$. Otherwise, for a $\SAB{i,j}$ test shift where neither $i$ or $j$ is a special leg, the term scales as $\Order{z^{\deg(i)+\deg(j)-4}}$. The degree of a leg refers to the number of propagators that connect to the leg in the tree diagram. The best fall off occurs when both leg $i$ and $j$ have only one connection, where the term scales as $\Order{z^{-2}}$. The tree formula and the $\N=7$ BCFW is complementary in the sense that both manifest the $\Order{z^{-2}}$ scaling term by term, but under different conditions of test shift legs.


\section{$\N=8$ bonus relations and $\N=7$ bonus scaling: the MHV case}
After demonstrating our proof, we would like to show that $\N=7$ BCFW terms manifest the improved scaling because they are using ``bonus relations", which come from the $\Order{z^{-2}}$ fall off of $\N=8$ amplitudes. The bonus scaling of $\N=8$ amplitudes enables us to multiply a linear function of $z$ on our amplitude and deform $z$ as in BCFW recursion, except that we do not have to consider the boundary integral. These extra relations are called ``bonus relations". Multiplying by the $s$ channel, we have the sum over residues at $z=z_k$,
\begin{equation}
s(0) \, \M^{\N=8}_n = \sum\limits_k s(z_k)\int\di^8\e_{\Ph} \, \M_L \, \frac{1}{P^2} \, \M_R .
\end{equation}

Our purpose is to use the bonus relations to recombine $\N=8$ terms and cancel out linear relations between terms, such that the remaining expression corresponds to the $\N=7$ representation. The following analysis focuses on the MHV case for simplicity and parallels Appendix C of~\cite{Nandan:2012rk}. Note that the BCFW representation for the $\N=8$ n-point MHV amplitude will always have one more diagram than $\N=7$. We will show that we can use the bonus relation to express the additional $\N=8$ term using terms appearing in $\N=7$. More explicitly, we write the $\N=8$ n-point MHV amplitude as $\M(123\cdots n)$ or $\M^{\N=8}_n$, the $\N=7$ amplitude as $\M(1^-2^-3^+\cdots n^+)$ or $\M^{\N=7}_n$, and construct the BCFW representation using the $\SAB{n^+,1^-}$ shift:
\begin{equation}
	\sk{\h{n}} = \sk{n} + w \sk{1} ,\quad
	\ak{\h{1}} = \ak{1} - w \ak{n} ,\quad
	\h{\e}_n = \e_n + w \, \e_1 .
\end{equation} 
The $\N=8$ representation has $n-2$ diagrams while the $\N=7$ representation has $n-3$ diagrams. The additional term for $\N=8$ can be written as
\begin{equation}
	\int\di^8\e_{\Ph} \, \M_L \, \frac{1}{P^2} \, \M_R(\h{1}\h{P}2).
\end{equation}
Intuitively, we want to expand this term into the other $n-3$ terms, so we separate the additional term and multiply $S_{12}$ on each side
\begin{align}
	\M^{\N=8}_n &= \int\di^8\e_{Ph} \, \M_L \, \frac{1}{P^2} \, \M_R(\hat{1}\hat{P}2) + \sum\limits_{k=3}^{n-1} \int\di^8\e_{\Ph} \, \M_L \, \frac{1}{P^2} \, \M_R(\h{1}\h{P}k) ,\\
	s_{12}(0) \, \M^{\N=8}_n &= \sum\limits_{k=3}^{n-1} s_{12}(z_k)\int\di^8\e_{\Ph} \, \M_L \, \frac{1}{P^2} \, \M_R(\h{1}\h{P}k) .
\end{align}
After some manipulation, we successfully expand the additional term in $\N=8$ using others terms which have correspondence with $\N=7$.
\begin{align}
	\label{as}
	\M^{\N=8}_n &= \sum\limits_{k=3}^{n-1} \frac{s_{12}(z_k)}{s_{12}(0)} \int\di^8\e_{\Ph} \,  \M_L \, \frac{1}{P^2} \, \M_R(\h{1}\h{P}k) .
\end{align}

To compare with $\N=7$, we need to reduce the $\N=8$ terms to $\N=7$. In the MHV case, legs 1 and 2 are in multiplet $\Phi^-$, which have helicity $+3/2$, while the other particles are in multiplet $\Phi^+$, which has helicity $+2$, so we integrate out $\e_1^8, \e_2^8$ and $\h{\e_P}$ in the integral in (\ref{as}) as follows:
\begin{align}
	\nonum & \int\di^8\e_{\Ph} \, \M_L \, \frac{1}{P^2} \, \M_R(\h{1}\h{P}k)
	\\
	\nonum &= \int\di\e_1^8\di\e_2^8 \int\di\h{\e}_P^8 \, \delta(\ak{n}\h{\e}_n^8+\ak{\h{P}}\e_{\Ph}^8+\cdots) \, \delta(\SB{1k}\e_{\Ph}^8+\SB{k\h{P}}\eta_1^8+\SB{\h{P} 1}\eta_k^8)
	\int\di^7\e_{\Ph} \,
	\wtd{\M}_L\,\frac{1}{P^2} \,\wtd{\M}_R
	\\
	\nonum &= (w\AB{2 n}\SB{1 k}+\SB{k\h{P}}\AB{2\h{P}})
	\int\di^7\e_{\Ph} \, \wtd{\M}_L\,\frac{1}{P^2} \,\wtd{\M}_R \\
	&= \AB{12}\SB{1k} \int\di^7\e_{\Ph} \, \wtd{\M}_L\,\frac{1}{P^2} \,\wtd{\M}_R .
\end{align}
where $\wtd{\M}_L$ and $\wtd{\M}_R$ are $\M_L$ and $\M_R$ with the supermomentum conservation delta function stripped off.
Combining this result with (\ref{as}), we obtain
\begin{align}
	\sum\limits_{k=3}^{n-1}\int\mathrm{d^7}\eta_{\hat{P}} \, \AB{\hat{1} 2}\SB{1 k} \, \wtd{\M}_L\,\frac{1}{P^2} \,\wtd{\M}_R .
\end{align}
which is exactly the explicit form for the corresponding $\N=7$ BCFW representation:
\begin{align}
\sum\limits_{k=3}^{n-1} \int\di^7\e_{\Ph}\, \M_L^{\N=7}\,\frac{1}{P^2}\,\M_R^{\N=7} = \sum\limits_{k=3}^{n-1} \int\di^7\e_{\Ph}\,\AB{2\h{P}}\SB{\hat{P} k} \wtd{\M}_L\frac{1}{P^2}\wtd{\M}_R .
\end{align}
What we have demonstrated is that we can use a bonus relation to relate $\N=8$ BCFW terms to $\N=7$ BCFW terms. In other words, the reason why $\N=7$ BCFW terms have nicer large $z$ behavior in this example is precisely because they are implicitly using bonus relations to cancel out linear dependent terms which appear in the $\N=8$ representation.

The next question we can ask is whether the result applies to the general n-point $\NkMHV$ case. To answer this question, we try the same analysis on the 6-point NMHV amplitude. Now we have 14 terms in $\N=8$ compared with 9 terms in $\N=7$, so we require 5 bonus relations to reduce the additional 5 terms to the other 9 terms. We cannot continue, since we only have one bonus relation and it is impossible to solve 5 parameters with one condition in general. This implies the $\Order{z^{-2}}$ large $z$ behavior of $\N=7$ individual terms include not only bonus relations which cancel out linear dependence but also some unknown property in $\N=7$.


\section{Bonus scaling of ``bad shift" BCFW for string amplitudes}

Applications of BCFW recursion to string amplitudes have demonstrated improved large $z$ scaling compared to field theory amplitudes in certain kinematic regimes~\cite{Boels:2008fc}~\cite{Boels:2010bv}. This not only validates the construction of a ``bad shift" recursion formula without the requirement of $\N=7$ supersymmetry, but also enables the application of our previous argument to pursue even better term-by-term large $z$ bonus scaling.

Since we encounter an infinite tower of massive states in string theory, we first demonstrate the validity of our argument in the case of a massive propagator. The previous derivation is modified such that the on-shell condition becomes $\Ph^2 = (P + w\,\sk{i}\ab{j})^2 = m^2$. The primary shift parameter $w$ and $\sk{\jh},\ak{\ih},\h{\e}_j,\Ph$ become:
\begin{equation} 
w_m = \frac{-P^2+m^2}{\ab{j}P\sk{i}}
\end{equation}
\begin{equation}
\sk{\jh}_m = \sk{j} + w_m \sk{i} ,\quad
\ak{\ih}_m = \ak{i} - w_m \ak{j} ,\quad
\h{\e_j}_m = \e_j + w_m \e_i ,
\end{equation}
\begin{equation} 
\Ph_m = \frac{P\ak{j}\sb{i}P-m^2\sk{i}\ab{j}}{\ASB{j|P|i}} \,.
\end{equation}
In the numerator of $w_m$, the additional $m^2$ term scales as $z^0$ while the original $P^2$ scales as $z$, so the large $z$ scaling of $w_m$ and hence $\sk{\jh},\ak{\ih},\h{\e}_j$ are not affected. The large $z$ scaling of $\Ph_m$ is $\Order{z^0}$, which is also unchanged compared to that of the massless $\Ph$. Hence making the propagator massive does not affect the large $z$ behavior under the $\SAB{i,j}$ test shift.

It was shown in~\cite{Boels:2010bv} that the large $z$ scaling under a $\SAB{i,j}$ shift of superstring gluon amplitudes is improved by $z^{-\alpha's_{ij}}$ compared to the corresponding field theory amplitude. For an $\SAB{j^+,i^-}$ adjacent bad shift, the superstring amplitude scales as $z^{-\alpha's_{ij}+3-\N}$ since the corresponding super Yang-Mills amplitude scales as $z^{3-\N}$, thus by requiring the amplitude fall off faster than $z^0$, this leads to the kinematic condition $\text{Re}\left[3-\N-\alpha's_{ij}\right] < 0$ for a valid representation. Following our previous result (\ref{scaling}), under a $\SAB{i^-,j^+}$ test shift the $\N$ bad shift representation has $z^{\N-5}$ term-by-term scaling, compared to the $z^{-\alpha's_{ij}-1}$ large $z$ fall off of the whole amplitude. Note the curious result that for $3-\N < \text{Re}\left[\alpha's_{ij}\right] < 4-\N$, the term-by-term scaling is actually better than the whole amplitude. We turn to a specific amplitude for further investigation.

As an example, we look at the superstring four-point gluon component amplitude, which is given by:
\begin{equation}
\label{superstring_4_pt}
A_4(1^-, 2^-, 3^+, 4^+) = \frac{\AB{12}^4}{\AB{12}\AB{23}\AB{34}\AB{41}} \frac{\Gamma(1+\alpha's)\Gamma(1+\alpha't)}{\Gamma(1+\alpha'(s+t))}
\end{equation}
Here the $s$ and $t$ are the usual Mandelstam variables, which in our convention read as $s=s_{12}=(p_1+p_2)^2$, $t=s_{23}=(p_2+p_3)^2$, and $u=s_{13}=(p_1+p_3)^2$. The kinematic constraint for a valid recursion for this amplitude $\text{Re}\left[3-\alpha't\right] < 0$ was first given in~\cite{Boels:2008fc} by demonstrating the vanishing of the boundary term. We construct a bad shift representation by first deforming the amplitude with a $\N=0$ $\SAB{3^+, 2^-}$ shift,
\begin{equation}
A_4(w) = \frac{(\AB{12}-w\AB{13})^3}{\AB{23}\AB{34}\AB{41}} \frac{\Gamma(1+\alpha's+w\alpha'\SB{12}\AB{13}) \Gamma(1+\alpha't)} {\Gamma(1+\alpha'(s+t)+w\alpha'\SB{12}\AB{13})} .
\end{equation}
From the asymptotic expansion of the ratio of gamma functions, which can be obtained by using Stirling's series,
\begin{equation}
\frac{\Gamma(z+\alpha)}{\Gamma(z+\beta)}=z^{\alpha-\beta}\left[1+\frac{(\alpha-\beta)(\alpha+\beta-1)}{2z}\Order{z^{-2}}\right] \,,
\end{equation}
we can readily see that $A_4(w)$ indeed scales as $w^{-\alpha't+3}$.

Using the function $\frac{A_4(w)}{z}$, we can form the $\SAB{3^+, 2^-}$ representation of the amplitude as the sum of the residues 
at $w=-\frac{k+\alpha's}{\alpha'\SB{12}\AB{13}}, k\in\mathbb{N}$. This representation can be simplified into
\begin{equation}
\label{bad_shift_representation}
A_4(1^-, 2^-, 3^+, 4^+)
=
\frac{\AB{12}^4}{\AB{12}\AB{23}\AB{34}\AB{41}}
\frac{-1}{\alpha'^3 s^3}
\sum_{k=1}^{\infty}
\binom{\alpha't}{k}
\frac{(-1)^k k^4}{k+\alpha's} .
\end{equation}
Through direct summation using Mathematica, we can observe the convergence of the bad shift representation (\ref{bad_shift_representation}) to the closed form of the amplitude (\ref{superstring_4_pt}) within the kinematic regime $\text{Re}\left[3-\alpha't\right] < 0$. Another way to look at the convergence of the series is through the alternating series test. The ratio between terms of the series $a_k$ expands at large $k$ as
\begin{equation}
r = \left|\frac{a_{k+1}}{a_k}\right| = 1 + \frac{3-\alpha't}{k} +\Order{k^{-2}} .
\end{equation}
We obtain the condition $3-\alpha't < 0$ by requiring $r<1$ for sufficiently large $k$ such that the series converges.

Under the $\SAB{2^-, 3^+}$ test shift, the $\SAB{3^+, 2^-}$ bad shift representation deforms into
\begin{equation}
A_4(z)
=
\frac{\AB{12}^4}{\AB{12}\AB{23}(\AB{34}+z\AB{24})\AB{41}}
\frac{-1}{\alpha'^3(s-z\AB{12}\SB{13})^3}
\sum_{k=1}^{\infty}
\binom{\alpha't}{k}
\frac{(-1)^k k^4}{k+\alpha'(s-z\AB{12}\SB{13})} .
\end{equation}
From this form, we can observe directly that individual terms of the series fall off as $z^{-5}$ as predicted. Also note that for $\alpha't = n \in \mathbb{N}$, the series terminates after $n$ terms and $A_4(z)$ has finite poles, in contrast to  the case for $\alpha't$ at generic values. This property can also be observed by shifting the closed form formula for $A_4$.

We now turn to the previously mentioned curiosity at $3 < \text{Re}[\alpha't] < 4$. Firstly, it is tested numerically by Mathematica that the series converges in this kinematic region and that under the $\SAB{2^-, 3^+}$ test shift, individual terms scale as $z^{-5}$ at large $z$, better than the $z^{-\alpha't-1}$ scaling of the amplitude in its closed form. We observe that the series converges slower at larger $z$, such that the number of terms required to sum to a certain fraction of the amplitude increases with $z$. From this, we expect that convergence issues may arise at the large $z$ limit, allowing the large $z$ fall off for individual terms to be better than the closed form in this kinematic region.

Similar analysis can be applied to the closed superstring. In our previous reasoning for supergravity, we noted that our argument for bonus scaling only applies to $\N=7$ since the amplitude scales as $z^{6-\N}$ under the bad shift, and thus only offers a valid representation for $\N>6$. For gravitons in the superstring, the condition for a valid $\SAB{j^+,i^-}$ ``bad shift" representation is:
\begin{equation}
\text{Re}\left[6 - \N - 2\alpha's_{ij}\right] < 0  .
\end{equation}
In this kinematic regime, the $\SAB{j^+,i^-}$ bad shift representation has $z^{\N-9}$ term-by-term large $z$ scaling under a $\SAB{i^-,j^+}$ test shift according to (\ref{scaling}), compared to the $z^{-2\alpha's_{ij}^2-2}$ scaling of the whole amplitude. Similarly, note that the term-by-term large $z$ fall off is better than the whole amplitude for $6- \N < \text{Re}\left[2\alpha's_{ij}\right] < 7- \N$.


\section{Conclusion and Future directions}

In this note, we prove that the  ``bad shift" BCFW representation of $\N=7$ supergravity gives building blocks that exhibit term by term bonus $\Order{z^{-2}}$ fall off. In particular, we prove that using the $\SAB{j^+,i^-}$ BCFW representation of $\NkMHV$ amplitudes, each term vanishes as $\Order{z^{-2}}$ under the $\SAB{i^-,j^+}$ deformation. Focusing on the MHV case, we find that the $\Order{z^{-2}}$ behavior is also present for a large number of other $\SAB{-,+}$ deformations. For example, in the $\SAB{n^+,1^-}$ representation, all $\SAB{2^-,m^+}$ deformation exhibits term by term $\Order{z^{-2}}$ asymptotic behavior. The reason that the ``bad shift" is a valid BCFW shift can be traced back to the $\Order{z^{-2}}$ fall off of $\N=8$ supergravity, which allows for the susy reduction to still have vanishing asymptotic, i.e. the shift behaves as $\Order{z^{-2}}$. Thus the ``bad shift" BCFW representation of $\N=7$ supergravity is the only BCFW recursion that utilizes the $\Order{z^{-2}}$ fall off of the amplitude. We demonstrate this claim by showing that for the MHV case, we can use the bonus relation to recombine building blocks in $\N=8$ BCFW into building blocks of the $\N=7$ bad shift. 

Our previous analysis only allows us to relate the BCFW representation of $\mathcal{N}=8$ supergravity to the  $\mathcal{N}=7$ bad shift representation for the MHV amplitude. This relation is no longer straightforward for NMHV amplitude and beyond. For example the six-point NMHV contains 14 diagrams in $\mathcal{N}=8$ supergravity versus 9 diagrams for $\mathcal{N}=7$ bad-shift representation. Since there is only one bonus relation at each multiplicity, it is insufficient to convert one representation to the other, unless one incorporates the information of the bonus relations for the lower point amplitudes. This would require us to further expand the BCFW representation. Indeed it is known that using all bonus relation, one can express the supergravity amplitudes in terms of $(n-3)!$ building blocks~\cite{He}. It will be interesting to see if one can utilize these building blocks to form term by term $\Order{z^{-2}}$ fall off for all deformations.

Recent studies~\cite{David} have shown how BCFW terms of gravitational amplitudes can pair into combinations with improved permutation invariance, such that leading $\Order{z^{-1}}$ pieces cancel and $\Order{z^{-2}}$ fall off is exposed.  However, it appears that to have $\Order{z^{-2}}$ fall off for all shifts, one eventually requires the combination of everything and end up with the full amplitude, which is similar to the $\N=7$ bad shift result. Thus it would appear that the improved fall off obtained by implementing partial permutation invariance can be similarly achieved without. It might be interesting to perform a general search of rational functions of spinor products that satisfies the correct helicity weight, mass dimension, at most simple poles and $\Order{z^{-2}}$ fall off for all shifts. These are very stringent constraints, and it is likely that the solution can serve as the true building blocks for the amplitude. 

Finally, we note that the ``bad shift'' BCFW recursion is also valid for string amplitudes under certain kinematic conditions. Unlike the story for the $\N=7$ theory, whose validity of the ``bad shift'' BCFW is attributed to the bonus fall off of $\N=8$ gravity, here the validity of the string amplitude representation is tied to its improved high-energy behavior. Due to the enhanced large $z$ scaling of string amplitudes, the restriction to the $\N=7$ representation is lifted and we can further reduce supersymmetry to expose better term-by-term large $z$ fall off compared to field theory. Furthermore, just as the bonus scaling of the $\N=7$ bad shift representation may be considered as the incorporation of $\N=8$ bonus relations, the improved behavior of BCFW terms of string amplitudes hint at possible relations inviting deeper investigation. It would be interesting to understand further, whether or not new symmetry or new amplitude relations emerge from this picture.


\section*{Acknowledgements}

We thank Yu-tin Huang for suggesting this problem and the thoughtful conversations and encouragements throughout its course. We would also like to thank Congkao Wen and Kasper Larsen for useful discussions. Jin-Yu Liu and En Shih are supported by National Science Council, Taiwan, R.O.C Grant Number 100-2628-M-002-012-MY4.


\appendix

\section{Derivation of $\hat{P}$}
Consider a $\SAB{j,i}$ BCFW representation:
\begin{equation} 
\M_n = \sum \int\di^7\e_{\Ph}  \ \M_L \frac{1}{P^2} \M_R \  |_{\Ph^2=m^2} \,,
\end{equation} 
\begin{equation} 
\sk{\jh} = \sk{j} + w \sk{i} ,\quad
\ak{\ih} = \ak{i} - w \ak{j} ,\quad
\h{\e}_j = \e_j + w \, \e_i \,.
\end{equation}
\begin{equation}
\Ph = P + w\,\sk{i}\ab{j} .
\end{equation}
We can evaluate $w$ using the on-shell condition $\Ph^2=m^2$.
\begin{align*} 
\Ph^2
&= (P + w\,\sk{i}\ab{j})^2 \\
&= P^2 + 2 P\cdot w\,\sk{i}\ab{j} \\
&= P^2 + w\, \ASB{j|P|i} = m^2 .
\end{align*}
Therefore,
\begin{empheq}[box=\widefbox]{align} 
w = \frac{-P^2+m^2}{\ASB{j|P|i}} .
\end{empheq} \\
Plugging the expression for $w$ into $\hat{P}$,
\begin{align*}
	\hat{P} &= P + \frac{(-P^2+m^2)}{\ASB{j|P|i}}\sk{i}\ab{j} \\
	&= \frac{\SAB{i|P|j}P -P^2\sk{i}\ab{j} +m^2\sk{i}\ab{j}} {\ASB{j|P|i}} \,.
\end{align*}
This can be simplified by invoking the Schouten identity as follows:
\begin{align*}
	\ASB{j|P|i} P_{a\dot{b}}
	&= j_{\dot{c}} P^{\dot{c}d} i_d P_{a\dot{b}} \\
	&= - P_{\dot{c}}^{\ d} i_d P_{a}^{\ \dot{c}}j_{\dot{b}}
	- P_{a\dot{c}}j^{\dot{c}} P_{\dot{b}}^{\ d}i_d \\
	&= P_{a\dot{c}} P^{\dot{c}d}i_d j_{\dot{b}}
	+ P_{a\dot{c}}j^{\dot{c}} P_{\dot{b}d}i^d .
\end{align*}
Using $P^a_{\ \dot{c}} P^{\dot{c}d} = P^2 \epsilon^{ad}$, we have
\begin{align*}
	\ab{j}P\sk{i}P
	&=P^2 \delta_a^{\ d} i_d j_{\dot{b}}
	+ P_{a\dot{c}}j^{\dot{c}} P_{\dot{b}d}i^d \\
	&= P^2 i_a j_{\dot{b}}
	+ P_{a\dot{c}}j^{\dot{c}} P_{\dot{b}d}i^d \\
	&= P^2\sk{i}\ab{j} + P\ak{j}\sb{i}P .
\end{align*}
We obtain for $\hat{P}$:
\begin{empheq}[box=\widefbox]{align}
	\hat{P} &= \frac{P\ak{j} \sb{i}P +m^2\sk{i}\ab{j}} {\ASB{j|P|i}} .
\end{empheq}


\end{document}